\documentclass[twoside]{article}
\usepackage{spec,qic_spec}

\begin{document}
\setlength{\textheight}{8.0truein}    

\runninghead{Robustness of QMA against witness noise}
            {Friederike Anna Dziemba}

\normalsize\textlineskip
\thispagestyle{empty}
\setcounter{page}{1}


\vspace{0.88truein}

\alphfootnote

\fpage{1}

\centerline{\bf
ROBUSTNESS OF QMA AGAINST WITNESS NOISE}
\vspace{0.37truein}
\centerline{\footnotesize
FRIEDERIKE ANNA DZIEMBA\footnote{friederike.dziemba@itp.uni-hannover.de}}
\vspace{0.015truein}
\centerline{\footnotesize\it Institut f\"ur Theoretische Physik, Leibniz Universit\"at Hannover}
\baselineskip=10pt
\centerline{\footnotesize\it Appelstr. 2, 30167 Hannover, Germany
}
\vspace{0.015truein}
\centerline{\footnotesize Dated: September 12, 2017}

\vspace{0.21truein}

\abstracts{
Using the tool of concatenated stabilizer coding, we prove that the complexity class $\QMA$ remains unchanged even if every witness qubit is disturbed by constant noise. This result may not only be relevant for physical implementations of verifying protocols but also attacking the relationship between the complexity classes $\QMA$, $\QCMA$ and $\BQP$, which can be reformulated in this unified framework of a verifying protocol receiving a disturbed witness. While $\QCMA$ and $\BQP$ are described by fully dephasing and depolarizing channels on the witness qubits, respectively, our result proves 
$\QMA$ to be robust against $27\%$ dephasing and $18\%$ depolarizing noise. 
}{}{}

\vspace{10pt}


\vspace{1pt}\textlineskip    
\section{Introduction}

The main hope for quantum computers is the efficient solving of computational problems beyond classical means. Efficiently quantum decidable problems form the complexity class $\BQP$. A common model of a $\BQP$-protocol is a quantum circuit polynomially sized in the length of the problem instance followed by a single qubit output measurement deciding the problem by accepting with high completeness probability or with low soundness probability. 

Yet from the field of quantum Hamiltonian complexity many physically relevant problems emerge that are not known to lie in the class $\BQP$, such as the local Hamiltonian problem, the density matrix consistiency problem, the non-identity check of quantum circuits or the non-isometry testing of quantum channels \cite{watrousReview, bookatz, aharonov, kempe, gharibian}. These problems are only known to lie in the broader class $\QMA$ of efficiently quantum verifiable problems. A problem $A=(A_\text{yes}, A_\text{no})$ is efficiently quantum verifiable iff for every $x\in A_\text{yes}$  there exists a quantum state, the so-called witness, that can make the verifier corresponding to a $\BQP$-protocol accept with high probability while for $x\in A_\text{no}$ every witness only leads to acceptance with low probability.

Though the complexity class $\QMA$ originates as a theoretical construction from complexity theory, physical implementations of specific 
protocol variants are conceivable.
In such physical implementations one naturally encounters noise processes. The tools of fault tolerance allow one to design efficient quantum circuits robust against constant qubit noise occuring at every computation step \cite{aharonovFaultTolerance}. 
In this paper we consider the robustness of  $\QMA$ against noise on the supplied witness using similar tools.
This idea 
motivates the following definition of a complexity class with disturbed witness:

\vspace{12pt}
\begin{definition}\label{def:noisyQMA}
Let $T$ be a single qubit quantum channel. 
A problem $A=(A_\text{yes}, A_\text{no})$ lies in the complexity class $\QMA_T(c,s)$ 
iff there exists a family $V=(V_x)_{x\in\{0,1\}^*}$ of polynomial-time generated quantum circuits acting on $n_\text{v}+n_\text{w}$ qubits, $n_\text{v},n_\text{w}\in\poly(|x|)$, such that
\begin{align*}
&\text{(Completeness case)}\quad
\forall x\in A_\text{yes} \exists \rho\in\mathcal{D}(\mathbb{C}^{2^{n_\text{w}}}):\\
&\quad\tr\Big(\Pi_\text{acc} V_x \big(\ket{0}\bra{0}^{\otimes n_\text{v}} \otimes T^{\otimes n_\text{w}}(\rho)\big)V_x^\dagger\Big)\ge c\\
&\text{(Soundness case)}\quad
\forall x\in A_\text{no} \forall \rho\in\mathcal{D}(\mathbb{C}^{2^{n_\text{w}}}):\\
&\quad\tr\Big(\Pi_\text{acc} V_x \big(\ket{0}\bra{0}^{\otimes n_\text{v}} \otimes \rho\big) V_x^\dagger\Big)\le s
\end{align*}
%
{\noindent with $\Pi_\text{acc}$ the operator projecting a designated output qubit onto the computational basis state $\ket{1}$.}
\end{definition}
\vspace{12pt}

\begin{definition}
We define $\QMA_T:=\QMA_T(2/3,1/3)$.
\end{definition}
\vspace{12pt}

For the rest of the paper the length of the input problem instance will always be denoted by the variable $n_\text{in}$, the number of verifier qubits by  $n_\text{v}\in\poly(n_\text{in})$ and the number of witness qubits by $n_\text{w}\in\poly(n_\text{in})$, while $\poly$ stands for the family of all polynomial-time computable and polynomially bounded functions $\mathbb{N}_0\rightarrow \mathbb{N}$.

For the identity channel $T=\Id$ definition \ref{def:noisyQMA} is equivalent to the definition of $\QMA(c,s)$. For $\QMA:=\QMA(2/3,1/3)$ the tool of amplification allows a large freedom in the choice of completeness and soundness parameters, showing that $\QMA=\QMA(c,s)$ for all polynomial-time computable functions $c$ and $s$ between $e^{-q}\le s, c\le 1-e^{-q}$ with gap $c-s\ge 1/q$ and $q\in\poly(n_\text{in})$. The simplest form of amplification is by parallelizing the original $\QMA$-protocol polynomially many times and adjusting the overall output qubit depending on the measurement distribution of the parallelized protocol copies \cite{kitaev, aharonov}.

Our definition of $\QMA_T$ also covers other important complexity classes such as $\BQP$ and $\QCMA$. While $\BQP$ corresponds to $\QMA_T$ with $T$ the fully depolarizing channel, the class $\QCMA$ is described by $\QMA_T$ with $T$ a quantum-classical channel such as the fully dephasing channel.
Notice that in the definition of $\QMA_T$ the channel is not applied to the witness in the soundness case while according to the usual definition of $\QCMA$ \cite{kitaev,watrousReview,aharonov} the witness is always classical. But indeed the class $\QCMA$ stays unchanged if the verifier is required to work correctly for all quantum witnesses in the soundness case since
the fully dephasing channel transmits classical states unchanged and can be simulated efficiently by the verifier on each witness qubit by its finite dimensional Stinespring dilation at the beginning of the protocol. Consequently $\QCMA$ is correctly described by $\QMA_T$ with $T$ the fully dephasing channel.

Even after decades of research it is still unkown if the subset relations
\begin{align*}
\BQP\subseteq\QCMA\subseteq\QMA
\end{align*}
are actually strict or if perhaps two or all three classes are equivalent.
One might gain some new insight into the subset relations 
by studying $\QMA_T$ with parameter-dependent channels like the partly dephasing channel and the partly depolarizing channel
\begin{align*}
T^\text{deph}_\epsilon(\rho)&=(1-\epsilon)\rho + \epsilon \frac{\rho + X\rho X}{2}\\
T^\text{depol}_\epsilon(\rho)&=(1-\epsilon)\rho + \epsilon \frac{\mathbb{I}}{2},
\end{align*}
which ``interpolate'' between the complexity classes $\QMA$ -- $\QCMA$  and $\QMA$ -- $\BQP$, respectively.

\begin{figure}[tbh]
\begin{minipage}[h]{0.45\linewidth}\centering
\quad\quad \includegraphics[width=0.77\textwidth]{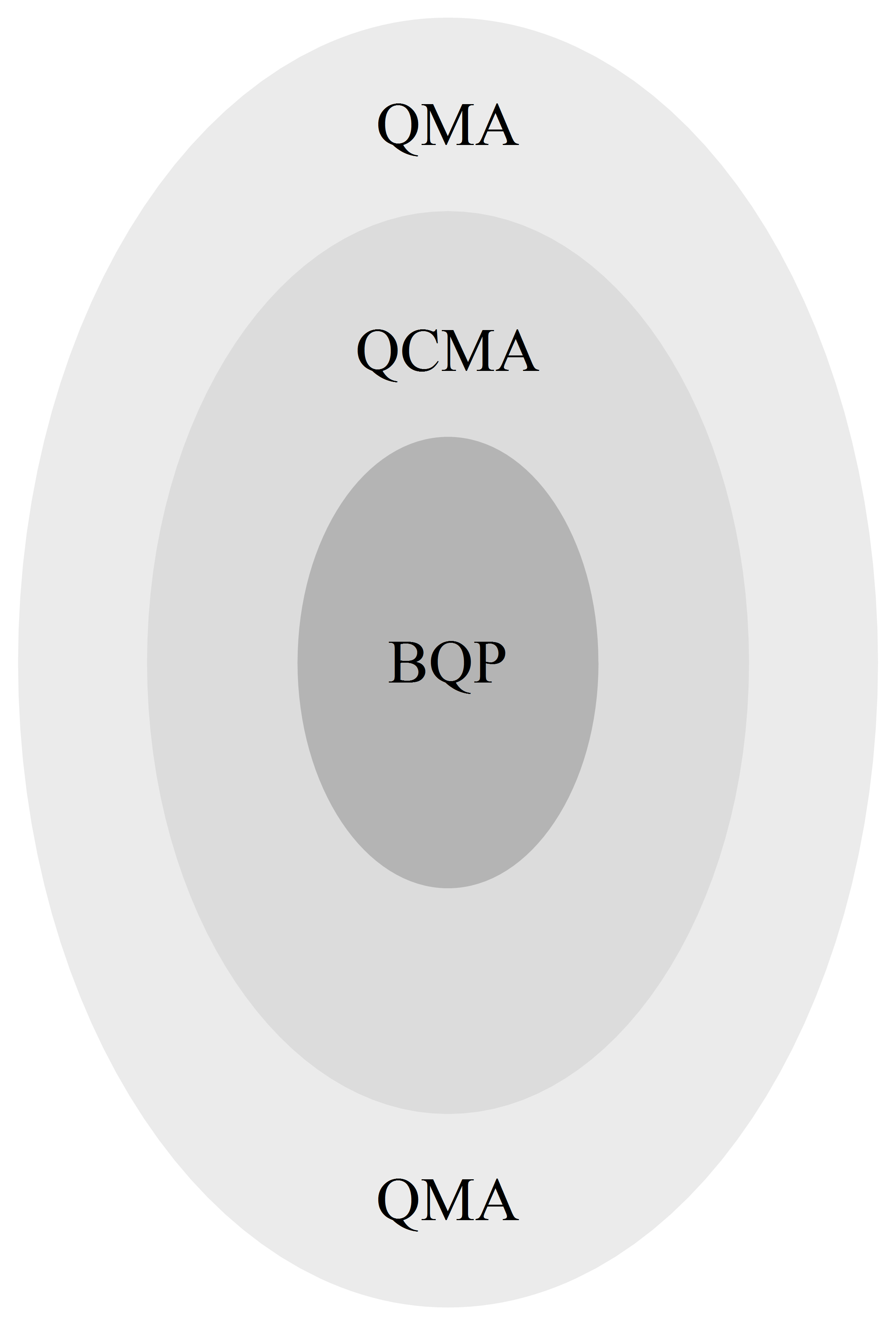}
\end{minipage}
\begin{minipage}[h]{0.02\linewidth}
\quad
\end{minipage}
\begin{minipage}[h]{0.45\linewidth}\centering
 \includegraphics[width=0.77\textwidth]{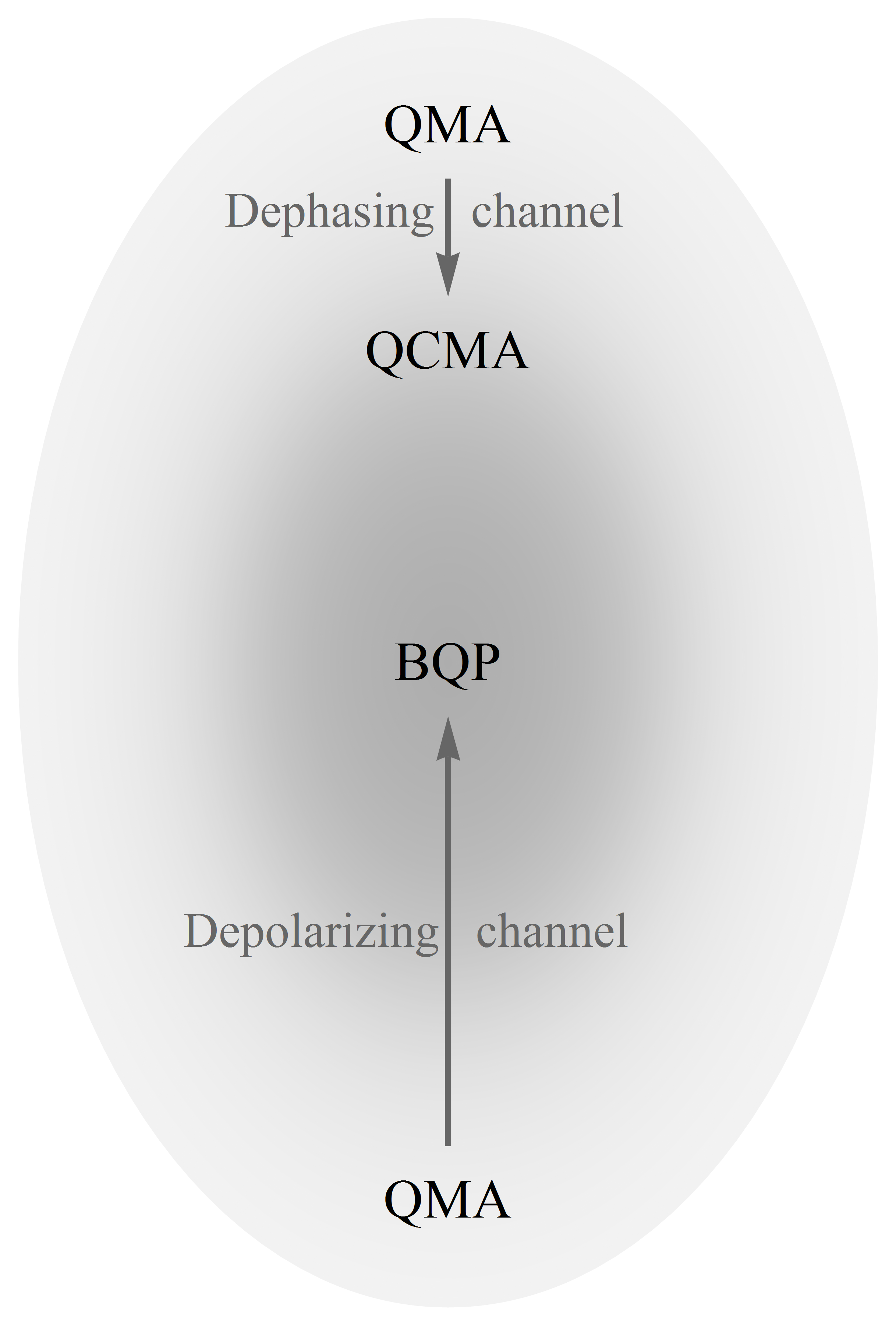}\quad\quad
\end{minipage}
\vspace{13pt}
\fcaption{\label{fig:disturbedWitnesses}Attacking $\BQP\subseteq\QCMA\subseteq\QMA$ by disturbed witnesses.}
\end{figure}

The authors of \cite{morimae} introduce a complexity variant of $\QMA$ with noisy witness as well, but with a different main intention. After briefly stating the existence of non-trivial channels that leave $\QMA$ invariant, the authors concentrate on proving a relaxed, error robust version of a graph state test to extend their statement to the $\QMA$ equivalent case in that the verifier is restricted to single qubit measurements.
In this paper we pursue proving in detail the robustness of $\QMA$ in its original form. While \cite{morimae} are satisfied with 
 at a pure existence statement, our 
constructive proof allows us to actually compute non-trivial upper bounds on the error parameter $\epsilon$ such that $\QMA_{T_\epsilon^\text{depol}}$ and $\QMA_{T_\epsilon^\text{deph}}$ still equal $\QMA$. We also compute a bound for general channels in terms of their deviation from the identity channel in diamond norm.

In contrast to \cite{morimae} we restrict our definition of a noisy $\QMA$ class to tensor product channels.
The assumption of independently and equally disturbed witness qubits 
is not only a physically relevant setting but also simplifies our constructive 
analysis while still covering our relevant cases $\QMA$, $\QCMA$ and $\BQP$.

In the next section we start with a simple approach by amplification that will prove robustness of $\QMA$ against qubit noise that decreases with the witness length. We obtain a much stronger result afterwards using the fault-tolerant tool of concatenated coding. The original $\QMA$-witness will be replaced by an encoded version passing the noise channel. The verifier exploits the redundance of coding to correct the noise and reconstruct the original witness as accurately as possible. This will prove the robustness of $\QMA$ even against constant qubit error:
\vspace{12pt}\noindent
\addtocounter{thmIntro}{+1}
\begin{thmIntro}\label{thm:central}
There is a constant $\epsilon_\Diamond>0$ such that $\QMA_T=\QMA$ for every quantum channel $T$ with $\Vert T-\Id\Vert_\Diamond\le\epsilon_\Diamond$.

Furthermore, for a channel $T$ of the above form, $\QMA_T(c,s)=\QMA_T$ for all polynomial-time computable functions $c$ and $s$ between $e^{-q}\le s, c\le 1-e^{-q}$ with gap $c-s\ge 1/q$ and $q\in\poly(n_\text{in})$.
\end{thmIntro}
\vspace{12pt}

For the proof of theorem \ref{thm:central} we will introduce the tool of concatenated coding in detail and derive the necessary performance properties in a straightforward way such that we are finally able to actually compute the tolerable error constant $\epsilon_\Diamond$ in the second-to-last-section. In the last section we will discuss some open questions.

\section{Robustness of $\QMA$ against decreasing qubit noise}

The goal of this section is to prove a first condition for channels $T$ that guarantees $\QMA_T=\QMA$.
Though not explicitly stated in definition \ref{def:noisyQMA} we allow the channel $T$ to depend on the witness length $n_\text{w}$ for this purpose. 
By simple norm inequalities and amplification by parallelization, we can prove that $\QMA$ stays unchanged as long as the disturbance on every witness decreases with the witness length:
\vspace{12pt}\noindent
\begin{thm}\label{thm:QMA_T=QMA}
$\QMA_T=\QMA$ for every constant $\delta>0$ and quantum channel $T$ with
\begin{align*}
\Vert T-\Id\Vert_\Diamond \le \frac{2(1-\delta)}{n_\text{w}}\text{.}
\end{align*}

Furthermore, for a channel $T$ of the above form, $\QMA_T(c,s)=\QMA_T$ for all polynomial-time computable functions $c$ and $s$ between $e^{-q}\le s, c\le 1-e^{-q}$ with gap $c-s\ge 1/q$ and $q\in\poly(n_\text{in})$.
\end{thm}
\vspace{12pt}\noindent
\begin{proof}
Since $\QMA_T\subseteq \QMA$ is trivial, we only have to prove the opposite subset relation. For this consider the $\QMA(1-e^{-n_\text{in}},e^{-n_\text{in}})$-verifier $V=(V_x)_{x\in\{0,1\}^*}$ for a problem $A$ as verifier of a $\QMA_T$-protocol. The soundness probability obviously remains unchanged.

To compute the completeness probability of the $\QMA_T$-protocol, we make use of the fact that the trace distance between two quantum states $\rho,\sigma$ equals
\begin{align*}
\Vert \rho-\sigma\Vert_1=2\max_{0\le\Lambda\le \mathbb{I}} \tr\left[\Lambda(\rho-\sigma)\right]
\end{align*}
\cite[lemma 9.1.1]{wilde}.
Notice that we use the same notation $\Vert\cdot\Vert_1$ for the trace norm of linear operators on a Hilbert space $\mathcal{H}$ as well as for the induced trace norm \cite{ahaKitMixed,watrousLecture} on superoperators $T\in L\big(L(\mathcal{H})\big)$:
\begin{align*}
\Vert T\Vert_1 := \sup_{\substack{\rho\in L(\mathcal{H})\\\rho\ne 0}} \frac{\Vert T(\rho)\Vert_1}{\Vert \rho\Vert_1}\text{.}
\end{align*}

The previous maximizing set includes the projection operator $\Lambda=V_x\Pi_\text{acc}V_x^\dagger$ for the final output measurement preceded by the verifier's circuit. Knowing this and some simple norm properties \cite{ahaKitMixed}, the difference in the acceptance probabilites supplied an undisturbed witness $\rho$ compared to the disturbed witness $T^{\otimes n_\text{w}}(\rho)$ can be estimated as follows:
\begin{align*}
\big\vert\tr\big[\Lambda \big(&\ket{0}\bra{0}^{\otimes n_\text{v}}\otimes(T^{\otimes n_\text{w}}(\rho)-\rho)\big)\big]\big\vert\\
&\le \frac{1}{2}\Vert \ket{0}\bra{0}^{\otimes n_\text{v}}\otimes(T^{\otimes n_\text{w}}(\rho)-\rho)\Vert_1\\
&\le\frac{1}{2} \Vert  \Id^{\otimes n_\text{v}} \otimes (T^{\otimes n_\text{w}}-\Id^{\otimes n_\text{w}})\Vert_1\\
&\le\frac{1}{2} \Vert  T^{\otimes n_\text{w}}-\Id^{\otimes n_\text{w}}\Vert_\Diamond\\
&=\frac{1}{2} \Vert (T\otimes \Id \otimes\Id\otimes\dots ) \circ (\Id \otimes T \otimes \Id \otimes \dots)\\
&\qquad\; \circ\dots - \Id^{\otimes n_\text{w}}\Vert_\Diamond\\
&\le\frac{1}{2} n_\text{w} \Vert  T-\Id\Vert_\Diamond\\
&\le 1-\delta\text{.}
\end{align*}

The completeness probability might hence decrease to $\delta-e^{-n_\text{in}}$.
Assume w.l.o.g.\ that $V$ outputs the correct answer deterministically for all inputs up to the length where $e^{-n_\text{in}} <\frac{\delta}{4}$ by hard encoding. Then $V$ is obviously a $\QMA_T(\frac{3\delta}{4},\frac{\delta}{4})$-verifier for the problem $A$. 
The statement $\QMA\subseteq \QMA_T(\frac{2}{3},\frac{1}{3})$ will be implied by amplification.

\vspace{1em}

Amplification is shown if we can construct a $\QMA_T(1-e^{-r},e^{-r})$-protocol for arbitrary $r\in\poly(n_\text{in})$  
and any problem $A\in\QMA_T(c,s)$ with $c$ and $s$ obeying the restrictions of the theorem. Such a construction is provided in a similar way to how $\QMA$ is amplified by parallelization \cite{kitaev,aharonov}. We only have to take the detour $\QMA_T(c,s)\subseteq\QMA(1-e^{-n_\text{in}},e^{-n_\text{in}})$ and make a sufficient $m$-fold parallelization of the existing $\QMA(1-e^{-n_\text{in}},e^{-n_\text{in}})$-verfier for $A$ that is a $\QMA_T(\frac{3\delta}{4},\frac{\delta}{4})$-verfier at the same time.

Define $\tilde{T}_{n_\text{w}}:=T_{m n_\text{w}}$ where the lower indexes denote the dependence parameter of the channels which we normally don't spell out. The sublety that each of the verifiers in parallel rather serves as $\QMA_{\tilde{T}}$-type than as $\QMA_T$-type protocol
due to the increased witness length is irrelevant since it still possesses a completeness probability of $\frac{3\delta}{4}$ and soundness probability of $\frac{\delta}{4}$ due to  
\begin{align*}
\Vert \tilde{T}-\Id\Vert_\Diamond &\le \frac{2(1-\delta)}{m n_\text{w}} \le \frac{2(1-\delta)}{n_\text{w}}
\end{align*}
and the same argumentation as before.
\end{proof}
\vspace{12pt}

The robustness statement so far is quite weak and qubit noise decreasing with the witness length is physically unrealistic. But we also just considered the scenario that the verifier receives the disturbed repetition of the witness that he would normally receive in the undisturbed case. Repetition is already a simple form of coding (though during amplification not explicitely used in the style of error correction and decoding). In the next sections we exploit the tool of concatenated stabilizer coding to prove that $\QMA$ is even robust against constant qubit noise. In this construction the verifier expects a disturbed encoded version of the original witness and is adjusted in the way that he first error corrects and decodes the received witness and then simulates the original $\QMA$-protocol.

\section{Concatenation of general codes}


Although nowadays the term  ``code'' is sometimes used in a wider sense, we speak of it according to the original concept of Knill and Laflamme \cite{knillLaflamme}, i.e., the encoding $\mathcal{E}$ is an isometric operation that is perfectly reversed by the decoding $\mathcal{D}$, a valid quantum channel and coisometry. The decoding is preceded by a recovery operation $\mathcal{R}$ whose purpose is to counter the noise that disturbs the encoded 
quantum state. 

The error correction theorem \cite[theorem 10.1]{nielsen} provides a neccessary and sufficient condition for a noise channel to be perfectly correctable by a given code. The constructive proof shows that in the case of correctibility it can always be achieved by a standard error correction channel consisting of a projective syndrome measurement followed by unitary recovery operators. The error correction condition furthermore just depends on the codespace projection but not on the specific encoding which is one reason why a code is often just defined as its codespace. 

But the specific encoding and recovery operation is relevant for the performance of a code against noise channels that it cannot correct perfectly. Since we use codes in such a manner, we always assume that a code $C$ comes along with a specific encoding $\mathcal{E}$, decoding $\mathcal{D}$ and standard error correction channel $\mathcal{R}$ with projective measurement $\{P_i\}_{i=1}^l$ and recovery operators $\{R_i\}_{i=1}^l$, i.e.,
\begin{align*}
\mathcal{R}(\rho)=\sum_{i=1}^l R_i P_i \rho P_i^\dagger R_i^\dagger\text{.}
\end{align*}

Moreover we only consider $[n,1]$ codes, i.e., those that map $1$ qubit to $n$ qubits, and product noise $T^{\otimes n}$ with $T$ a single qubit quantum channel. The goal of this section is to derive the expression for the \emph{coding map} of such a code that maps the noise channel $T$ to the \emph{effective noise channel} obtained by embracing the noise by encoding and recovery and decoding:
\vspace{12pt}
\begin{definition}
The \emph{coding map} $\Omega^C$ for a code $C$ is defined via
\begin{align*}
\Omega^C&: L\big(L(\mathbb{C}^2)\big) \rightarrow L\big(L(\mathbb{C}^2)\big)\\
 \Omega^C&(T)= \mathcal{D}\circ \mathcal{R} \circ T^{\otimes n} \circ \mathcal{E}\text{.}
\end{align*}
\end{definition}
\vspace{12pt}

Having reduced the noise by coding, we may regard the effective channel as new noise and approach the identity channel even further by applying another coding level. This is the idea of \emph{concatenated coding}. On the mathematical side we just have to concatenate the coding map $k$-times by itself to obtain the effective channel after $k$ coding levels and study if the series converges towards the identity channel.

\begin{figure}[t]
\centering
\begin{tikzpicture}[scale=0.5]
\draw (3.5,0.5) to (-1,0.5);
\filldraw[draw=black, fill=white] (0,0) rectangle (1,1) node[midway] {$T$};
\draw (3.5,1.7) to (-1,1.7);
\filldraw[draw=black, fill=white] (0,1.2) rectangle (1,2.2) node[midway] {$T$};
\draw (3.5,3.2) to (-1,3.2);
\filldraw[draw=black, fill=white] (0,2.7) rectangle (1,3.7) node[midway] {$T$};
\draw (3.5,4.4) to (-1,4.4);
\filldraw[draw=black, fill=white] (0,3.9) rectangle (1,4.9) node[midway] {$T$};
\draw (3.5,5.9) to (-1,5.9);
\filldraw[draw=black, fill=white] (0,5.4) rectangle (1,6.4) node[midway] {$T$};
\draw (3.5,7.1) to (-1,7.1);
\filldraw[draw=black, fill=white] (0,6.6) rectangle (1,7.6) node[midway] {$T$};
\draw (3.5,8.6) to (-1,8.6);
\filldraw[draw=black, fill=white] (0,8.1) rectangle (1,9.1) node[midway] {$T$};
\draw (3.5,9.8) to (-1,9.8);
\filldraw[draw=black, fill=white] (0,9.3) rectangle (1,10.3) node[midway] {$T$};
\draw (-2,0.25) rectangle (-1,1.95) node[midway] {$\mathcal{E}$};
\draw (-2,1.1) to (-3,1.1);
\draw (-2,2.95) rectangle (-1,4.65) node[midway] {$\mathcal{E}$};
\draw (-2,3.8) to (-3,3.8);
\draw (-2,5.65) rectangle (-1,7.35) node[midway] {$\mathcal{E}$};
\draw (-2,6.5) to (-3,6.5);
\draw (-2,8.35) rectangle (-1,10.1) node[midway] {$\mathcal{E}$};
\draw (-2,9.2) to (-3,9.2);
\draw (-4,0.6) rectangle (-3,4.3) node[midway] {$\mathcal{E}$};
\draw (-4,2.45) to (-5,2.45);
\draw (-4,6) rectangle (-3,9.7) node[midway] {$\mathcal{E}$};
\draw (-4,7.85) to (-5,7.85);
\draw (-6,1.45) rectangle (-5,8.85) node[midway] {$\mathcal{E}$};
\draw (-6,5.15) to (-7.2,5.15);
\node at (-7.7,5.15) {\small $\dots$};
\draw (4.5,1.1) to (7,1.1);
\filldraw[draw=black, fill=white] (3,0.25) rectangle (2,1.95) node[midway] {$\mathcal{R}$};
\filldraw[draw=black, fill=white] (4.5,0.25) rectangle (3.5,1.95) node[midway] {$\mathcal{D}$};
\draw (4.5,3.8) to (7,3.8);
\filldraw[draw=black, fill=white] (3,2.95) rectangle (2,4.65) node[midway] {$\mathcal{R}$};
\filldraw[draw=black, fill=white] (4.5,2.95) rectangle (3.5,4.65) node[midway] {$\mathcal{D}$};
\draw (4.5,6.5) to (7,6.5);
\filldraw[draw=black, fill=white] (3,5.65) rectangle (2,7.35) node[midway] {$\mathcal{R}$};
\filldraw[draw=black, fill=white] (4.5,5.65) rectangle (3.5,7.35) node[midway] {$\mathcal{D}$};
\draw (4.5,9.2) to (7,9.2);
\filldraw[draw=black, fill=white] (3,8.35) rectangle (2,10.1) node[midway] {$\mathcal{R}$};
\filldraw[draw=black, fill=white] (4.5,8.35) rectangle (3.5,10.1) node[midway] {$\mathcal{D}$};
\draw (8,2.45) to (10.5,2.45);
\filldraw[draw=black, fill=white] (6.5,0.6) rectangle (5.5,4.3) node[midway] {$\mathcal{R}$};
\filldraw[draw=black, fill=white] (8,0.6) rectangle (7,4.3) node[midway] {$\mathcal{D}$};
\draw (8,7.85) to (10.5,7.85);
\filldraw[draw=black, fill=white] (6.5,6) rectangle (5.5,9.7) node[midway] {$\mathcal{R}$};
\filldraw[draw=black, fill=white] (8,6) rectangle (7,9.7) node[midway] {$\mathcal{D}$};
\draw (11.5,5.15) to (12.7,5.15);
\node at (13.3,5.15) {\small $\dots$};
\filldraw[draw=black, fill=white] (10,1.45) rectangle (9,8.85) node[midway] {$\mathcal{R}$};
\filldraw[draw=black, fill=white] (11.5,1.45) rectangle (10.5,8.85) node[midway] {$\mathcal{D}$};
\draw[dashed] (-2.5,-1) rectangle (5,2.45);
\draw[dashed] (-4.5,-1.5) rectangle (8.5,5.15);
\draw[dashed] (-6.5,-2) rectangle (12,10.55);
\node at (12.7,-2.4) {\emph{$\ddots$}};
\end{tikzpicture}
\fcaption{3-fold concatenation of a $[2,1]$-code.}
\end{figure}

Notice that we can regard the effective channel as new noise for another coding level only because our initial assumption of product noise ensures that the overall noise after $k$ coding levels is given by $T^{\otimes n^k}$. If this assumption was dropped, we would need knowledge about how the noise extends to larger systems and could not simply reduce our study to the fixed point analysis of the coding map $\Omega^C$.

Since complete positivity and trace-preservation are irrelevant for the analysis of a coding map, its domain is extended to all superoperators.
Due to the linearity of superoperators, it is often useful to consider their effect on arbitrary linear operators. That is also why the notation of the following convention will be used for arbitrary linear single qubit operators $\rho_0$ and not just for valid density operators.

\vspace{12pt}
\begin{con}\label{con:codeProtocol}
We use the following notation for the different stages an initial operator $\rho_0\in L(\mathbb{C}^2)$ runs through while passing one level of coding:
\begin{align*}
\rho_0^C=\mathcal{E}(\rho_0);\;\;
\rho^C=T^{\otimes n}(\rho_0^C);\;\;
\rho_f^C=\mathcal{R}(\rho^C);\;\;
\rho_f=\mathcal{D}(\rho_f^C)
\end{align*}
\begin{minipage}{\textwidth}
\centering
\begin{tikzpicture}[scale=0.4]
\draw (-1.5,1.5) to (1.5,1.5);
\filldraw[draw=black, fill=white] (-0.5,1) rectangle (0.5,2) node[midway] {$T$};
\draw (-1.5,0) to (1.5,0);
\filldraw[draw=black, fill=white] (-0.5,-0.5) rectangle (0.5,0.5) node[midway] {$T$};
\draw (-1.5,-1.5) to (1.5,-1.5);
\filldraw[draw=black, fill=white] (-0.5,-2) rectangle (0.5,-1) node[midway] {$T$};
\draw (-5.5,1.5) to (-3.5,1.5);
\draw (-6.5,0) to (-3.5,0);
\draw (-5.5,-1.5) to (-3.5,-1.5);
\filldraw[draw=black, fill=white] (-5.5,-2) rectangle (-4.5,2) node[midway] {$\mathcal{E}$};
\draw (3.5,1.5) to (6.5,1.5);
\draw (3.5,0) to (6.5,0);
\draw (3.5,-1.5) to (6.5,-1.5);
\filldraw[draw=black, fill=white] (4.5,-2) rectangle (5.5,2) node[midway] {$\mathcal{R}$};
\draw (8.5,1.5) to (10.5,1.5);
\draw (8.5,0) to (11.5,0);
\draw (8.5,-1.5) to (10.5,-1.5);
\filldraw[draw=black, fill=white] (9.5,-2) rectangle (10.5,2) node[midway] {$\mathcal{D}$};
\node at (-7.5,0) {$\rho_0$};
\node at (-2.5,0) {$\rho_0^C$};
\node at (2.5,0) {$\rho^C$};
\node at (7.5,0) {$\rho_f^C$};
\node at (12.5,0) {$\rho_f$};
\end{tikzpicture}
\end{minipage}
\vspace{0.01em}

Given the initial and final operator we denote the expectation values of Pauli operators $\sigma\in\mathcal{P}=\{\mathbb{I},X,Y,Z\}$ by $\langle \sigma \rangle_0 = \tr(\sigma \rho_0)$ and $\langle \sigma \rangle_f = \tr(\sigma \rho_f)$, respectively.
\end{con}
\vspace{12pt}

Since every operator in $L(\mathbb{C}^2)$ can be written as a linear combination of Pauli operators, a superoperator $T\in L\big(L(\mathbb{C}^2)\big)$ is fully described by its \emph{Stokes parametrization}: a complex $4\times 4$ matrix with the matrix entry $T_{\sigma\sigma'}$ representing the prefactor of $\sigma\in \mathcal{P}=\{\mathbb{I},X,Y,Z\}$ in the output operator given the input operator $\sigma'\in \mathcal{P}$. Observe that the prefactor of $\sigma$, given the linear combination $\rho\in L(\mathbb{C}^2)$, equals exactly half of the expectation value of $\sigma$:
\begin{align*}
\rho&=\frac{1}{2} \sum_{\sigma\in\mathcal{P}} \langle \sigma \rangle \sigma\text{.}
\end{align*}
Consequently the matrix element $T_{\sigma \sigma'}$ can also be considered as the expectation value of the Pauli operator $\sigma$ given the operator $T\left(\frac{1}{2}\sigma'\right)$:
\begin{align*}
T_{\sigma \sigma'}=\tr\left(\sigma T\left(\frac{1}{2}\sigma'\right)\right)\text{.}
\end{align*}



Before deriving an expression for the coding map, let us first recall some properties of the Stokes representation from \cite{ruskaiDelta,fern}. First of all the matrix entries of a valid quantum channel are real-valued and lie within the interval ${[-1,1]}$. Furthermore the first row equals $(1,0,0,0)$ iff the superoperator is trace-preserving. Trace-preserving superoperators that are \emph{diagonal} in the Stokes representation play a special role in the following analysis and will be fully described by the notation $T=[T_{XX}, T_{YY}, T_{ZZ}]$. The following correspondance is easy to check:

\vspace{12pt}
\begin{lem}\label{lem:pauliChannel}
Superoperators of the form
\begin{align*}
T(\rho)=p_\mathbb{I}\rho + p_X X\rho X + p_Y Y \rho Y + p_Z Z\rho Z
\end{align*}
are exactly those that are diagonal in Stokes representation. For trace-preserving superoperators ($p_\mathbb{I}=1-p_X-p_Y-p_Z$) the corresponding Stokes representation is
\begin{align*}
T=[1-2(p_Y + p_Z),1-2(p_X + p_Z),1-2(p_X + p_Y)]\text{.}
\end{align*}
\end{lem}
\vspace{12pt}

The $T$ of the above form is a valid quantum channel, namely a so-called \emph{Pauli channel}, iff $p_X,p_Y,p_Z\in{[0,1]}$ and $p_X+p_Y+p_Z\le 1$.
Notice that this implies that the point $(T_{XX},T_{YY},T_{ZZ})$ specified by a valid diagonal quantum channel $T$ always lies within the tetrahedron defined by the corners $(1,1,1)$, $(1,-1,-1)$, $(-1,1,-1)$ and $(-1,-1,1)$.

\vspace{1em}

We will be able to derive an expression for the coding map with the help of two simple expressions relating the initial and final expectation values $\langle \sigma\rangle_0$ and $\langle \sigma\rangle_f$ to the so-called en- and decoding operators of the code.  One might find the terminology ``decoding operators'' a bit misleading since the operators describe the action of the error correction channel rather than the actual decoding. However, we stick to this terminology to stay consistent with the literature \cite{rahn} from which we adapted the following definitions and proposition.

The following definition of the encoding operators does not rely explicitly to the encoding operation $\mathcal{E}$ but instead on the implied \emph{logical} Pauli operators. For every Pauli operator $\sigma\in\mathcal{P}$ we denote its logical counterpart by $\bar{\sigma}$ which is defined such that it acts on an encoded state like $\sigma$ on the original state, i.e. for all $\rho\in L(\mathbb{C}^2)$ it holds
\begin{align*}
\bar{\sigma}\mathcal{E}(\rho) \bar{\sigma}^\dagger = \mathcal{E}(\sigma \rho \sigma^\dagger)\text{.}
\end{align*}

\vspace{12pt}
\begin{definition}\label{def:encodingOp}
The \emph{encoding operators} $E_\sigma$, $\sigma\in\mathcal{P}$, for a quantum code $C$ are defined via
\begin{align*}
E_\sigma=\frac{1}{2}P_C \bar{\sigma}
\end{align*}
with $P_C$ the projection onto the codespace.
\end{definition}
\vspace{12pt}
Since encoding operators act like one half times the respective logical Pauli operators on the codespace and vanish on its orthogonal complement, an encoded state can be expressed as follows:
\vspace{12pt}
\begin{cor}\label{cor:encodingOp}
A quantum code $C$ encodes an initial operator $\rho_0$ into
\begin{align*}
\rho^C_0= \sum_{\sigma\in\mathcal{P}} \langle\sigma\rangle_0 E_\sigma\text{.}
\end{align*}
\end{cor}
\vspace{12pt}
\begin{definition}\label{def:decodingOp}
The \emph{decoding operators} $D_\sigma$, $\sigma\in\mathcal{P}$, of a quantum code $C$ are defined via
\begin{align*}
D_\sigma=2\sum_{j=1}^l P_j^\dagger R_j^\dagger E_\sigma R_j P_j\text{.}
\end{align*}
\end{definition}
\vspace{12pt}
\begin{cor}\label{cor:decodingOp}
The Pauli expectation values of the output operator $\rho_f$ for a quantum code $C$ are given by
\begin{align*}
\langle\sigma\rangle_f =\tr\left(D_\sigma \rho^C\right)\text{.}
\end{align*}
\end{cor}
\vspace{12pt}
\begin{prop}\label{prop:codingMap}
Consider a quantum code $C$ with encoding and decoding operators
\begin{align*}
E_{\sigma'}&= \sum_{\mu\in\mathcal{P}^{\otimes n}} \alpha_{\mu}^{\sigma'} \left(\frac{1}{2} \mu_1\right) \otimes \dots \otimes \left(\frac{1}{2} \mu_n\right)\\
D_\sigma&= \sum_{\nu\in\mathcal{P}^{\otimes n}} \beta_{\nu}^{\sigma} \nu_1 \otimes \dots \otimes \nu_n\text{.}
\end{align*}
Then a coding level transforms a single qubit superoperator $T$ into to the effective superoperator $\tilde{T}=\Omega^C(T)$ with
\begin{align*}
\tilde{T}_{\sigma\sigma'} = \sum_{\mu,\nu\in\mathcal{P}^{\otimes n}} \beta_{\nu}^\sigma \alpha_{\mu}^{\sigma'} \prod_{j=1}^n T_{\nu_j \mu_j}\text{.}
\end{align*}
\end{prop}
\vspace{12pt}
\noindent
\begin{proof}
The matrix element $\tilde{T}_{\sigma\sigma'}$ equals the final expectation value $\langle\sigma\rangle_f$ given the initial operator $\rho_0=\frac{1}{2}\sigma'$. Hence we obtain:
\begin{align*}
\tilde{T}_{\sigma\sigma'}&=\langle \sigma\rangle_f\\
&
{=}\tr(D_\sigma \rho^C)\\
&=\tr\left(D_\sigma T^{\otimes n}(\rho_0^C)\right)\\
&
{=}\tr\left(D_\sigma T^{\otimes n}(E_{\sigma'})\right)\\
&=  \sum_{\mu,\nu\in\mathcal{P}^{\otimes n}} \beta_{\nu}^\sigma \alpha_{\mu}^{\sigma'} \prod_{j=1}^n \tr\left( \nu_j T\left(\frac{1}{2}\mu_j\right)\right)\\
&= \sum_{\mu,\nu\in\mathcal{P}^{\otimes n}} \beta_{\nu}^\sigma \alpha_{\mu}^{\sigma'} \prod_{j=1}^n T_{\nu_j \mu_j}\text{.} \qedhere
\end{align*}
\end{proof}
\vspace{12pt}

The previous proposition shows that the matrix elements of a coding map are polynomials of degree $n$ in the matrix entries of the input superoperator. A smaller physical qubit number $n$ might hence simplify the analysis of a concatenated quantum code. However, in many cases a certain structure preservation is even more important for simplification. Concatenated stabilizer codes provide for example the advantage to be diagonality preserving as we will see in the next section.
\section{Concatenation of stabilizer codes}


In this section we want to show the derivation of the coding map for stabilizer codes as stated in \cite{rahn}. Preceding this we give a short review on stabilizer codes and lay out the notation we use in the following sections.

The codespace of a stabilizer code is the common $+1$-eigenspace of commuting stabilizers which form a subgroup of the Pauli group
\begin{align*}
\mathcal{P}_n:=\left\{ c\cdot \sigma \,|\, c\in \{\pm 1,\pm i\}, \sigma\in \mathcal{P}^{\otimes n}\right\}
\end{align*}
with $\mathcal{P}=\{\mathbb{I},X,Y,Z\}$ the set of Pauli operators.
We prohibit $-\mathbb{I}$ to be an element of a valid stabilizer group since its common $+1$-eigenspace is obviously empty. This condition guarantees all stabilizers to be hermitian.

A stabilizer group of an $[n,k]$ stabilizer code can always be generated by $m=n-k$ independent generators. We will continue assuming $k=1$ and denote the generators of a stabilizer code by $\{g_i\}_{i=1}^m$ and its stabilizer group by $S=\{S_i\}_{i=1}^{2^m}$.


Stabilizer codes form the most famous class of quantum codes since they provide a strong algebraic structure based on the simplicity of Pauli operators that allows a simple and straightforward analysis. The crucial point is that the standard error correction channel constructed in the proof of the error correction theorem 
for a noise channel with operation elements $\{E_i\}_{i=1}^l$ also corrects any noise channel whose operation elements are linear combinations of the operators $\{E_i\}_{i=1}^l$. Since every linear operator on a qubit can be written as a linear combination of Pauli operators, the whole error correction analysis can be reduced to the consideration of Pauli errors. The error correction theorem phrased for stabilizer codes \cite[theorem 10.8]{nielsen} states that an error set $\{E_i\}_{i=1}^l\subseteq\mathcal{P}_n$ is correctable (i.e., any noise channel whose operation elements are linear combinations of the $\{E_i\}_{i=1}^l$ can be corrected) iff all pairs of errors $E_i$ and $E_j$ fulfill
\begin{align*}
E_i^\dagger E_j \notin N(S)\backslash S^\pm
\end{align*}
with $S^\pm$ the set of all stabilizers times prefactors $\{\pm 1,\pm i\}$ and $N(S)\subseteq \mathcal{P}_n$ the \emph{normalizer} of $S$ consisting of all elements that commute with the stabilizers.

The projection operators $\{P_i\}_{i=1}^{2^m}$ of the standard error correction channel furthermore turn out to equal the projection onto the $2^m$ orthogonal generator syndrome spaces while the recovery operators $\{R_i\}_{i=1}^{2^m}$ are Pauli group elements turning the syndrome spaces into the codespace.

Besides the stabilizers the normalizer contains the \emph{logical} Pauli operators of the code, denoted by $\bar{\sigma}$, that act on the encoded state like the Pauli operators $\sigma\in\mathcal{P}$ on the original state. Notice that the definition of the logical operators fixes the encoding of the code and vice versa. Since a logical operator maps a valid codespace state just into another codespace state, it is conceivable why two errors  differing by it cannot be distinguished by an error correction procedure.
A stabilizer code can correct up to $\lfloor \frac{d-1}{2}\rfloor$ arbitrary single qubit errors with $d$ the \emph{distance} of the code which is defined by the minimum Pauli weight over all elements of $N(S)\backslash S^\pm$. 

We call a stabilizer code an $[n,k,d,w]$ code if it is an $[n,k]$ code of distance $d$ and 
$w$ is the minimum Pauli weight over all non-identity stabilizer elements. This last parameter is relevant for the later convergence study which relies on $[n,1,d,w]$ codes with distance $d\ge 3$ and $w\ge 2$. Notice that such codes exist, e.g., the 5-qubit code, the Steane 7-qubit code and the Shor code, as listed in table \ref{tab:codes} have these parameters.

This short overview of stabilizer codes should suffice for our purposes. For more details see standard textbooks, such as \cite{nielsen}, or \cite{gottesmanThesis}.

\vspace{4pt}   
\begin{table}[hb]\label{tab:codes}
\tcaption{Common stabilizer codes.}
\begin{tabular}{l c c c}\\
\hline
{} &5-qubit code &Steane 7-qubit code &Shor code\\
\hline
Generators &$\X\Z\Z\X\I$ &$\I\I\I\X\X\X\X$ &$\Z\Z\I\I\I\I\I\I\I$\\
{} &$\I\X\Z\Z\X$ &$\I\X\X\I\I\X\X$ &$\I\Z\Z\I\I\I\I\I\I$\\
{} &$\X\I\X\Z\Z$ &$\X\I\X\I\X\I\X$ &$\I\I\I\Z\Z\I\I\I\I$\\
{} &$\Z\X\I\X\Z$ &$\I\I\I\Z\Z\Z\Z$ &$\I\I\I\I\Z\Z\I\I\I$\\
{} &\qquad &$\I\Z\Z\I\I\Z\Z$ &$\I\I\I\I\I\I\Z\Z\I$\\
{} &\qquad &$\Z\I\Z\I\Z\I\Z$ &$\I\I\I\I\I\I\I\Z\Z$\\
{} & &&$\X\X\X\X\X\X\I\I\I$\\
{} & &&$\I\I\I\X\X\X\X\X\X$\\
$\bar{\X}$ &$\X\X\X\X\X$ &$\X\X\X\X\X\X\X$ &$\X\X\X\X\X\X\X\X\X$\\
$\bar{\Z}$ &$\Z\Z\Z\Z\Z$ &$\Z\Z\Z\Z\Z\Z\Z$ &$\Z\Z\Z\Z\Z\Z\Z\Z\Z$\\
Recovery operators &$\mathrm{\mathbb{I},X_i, Y_i, Z_i}$ &$\mathrm{\mathbb{I},X_i,Z_j,X_iZ_j}$ &$\mathrm{\{\mathbb{I},X_1,X_2,X_3\}\cdot\{\mathbb{I},X_4,X_5,X_6\}}$\\
\qquad &$1\le i\le 5$ &$1\le i\le 7$ &$\mathrm{\cdot\{\mathbb{I},X_7,X_8,X_9\}}$\\
&&&$\mathrm{\cdot\{\mathbb{I},Z_1Z_2Z_3,Z_4Z_5Z_6,Z_7Z_8Z_9\}}$\\
\hline\\
\end{tabular}
\end{table}

\vspace{1em}

The goal of this section is to adapt the expression of the coding map provided by proposition \ref{prop:codingMap} for stabilizer codes under the restriction of diagonal, trace-preserving superoperators.
 For this we first introduce a shortening notation:
\vspace{12pt}
\begin{definition}
For a stabilizer code we define the \emph{f-function}
\begin{align*}
f_{i\sigma} = \sum_j \eta(R_j,S_i) \eta(R_j,\bar{\sigma})
\end{align*}
with
\begin{align*}
\eta(\mu,\nu) =
\begin{cases}
+1 &\quad\text{ if $\mu$ and $\nu$ commute}\\
-1 &\quad\text{ if $\mu$ and $\nu$ anticommute}
\end{cases}
\end{align*}
for $\mu,\nu\in\mathcal{P}_n$.
\end{definition}
\vspace{12pt}
\begin{lem}
The decoding operators of a stabilizer code can be written as
\begin{align*}
D_\sigma = \frac{1}{|S|} \sum_i f_{i\sigma} S_i\bar{\sigma}\text{.}
\end{align*}
\end{lem}
\vspace{12pt}
\begin{proof}
Since the codespace projection for stabilizer codes equals $P_C=\frac{1}{|S|}\sum_i S_i$ we can insert
\begin{align*}
E_\sigma= \frac{1}{2|S|}\sum_i S_i\bar{\sigma}
\end{align*}
into the expression for the decoding operators given by definition \ref{def:decodingOp}:
\begin{align*}
D_\sigma &= \frac{1}{|S|} \sum_i \sum_j P_j R_j^\dagger S_i\bar{\sigma} R_j P_j\\
&=\frac{1}{|S|} \sum_i \sum_j P_j \eta(R_j,S_i)\eta(R_j,\bar{\sigma}) S_i \bar{\sigma} P_j\text{.}
\end{align*}
Since the syndrome space projections $P_j$ equal $\frac{1}{|S|} \sum_i \eta(R_j, S_i) S_i$ and commute with all stabilizers and logical operators, we obtain
\begin{align*}
D_\sigma &=\frac{1}{|S|} \sum_i \sum_j \eta(R_j,S_i)\eta(R_j,\bar{\sigma}) S_i \bar{\sigma}\\
&= \frac{1}{|S|} \sum_i f_{i\sigma} S_i \bar{\sigma}\text{.}\qedhere
\end{align*}
\end{proof}
\vspace{12pt}
\begin{lem}\label{lem:coefficients}
The coefficients of the en- and decoding operators of a stabilizer code, as introduced in proposition \ref{prop:codingMap}, fulfill
\begin{align*}
\alpha_{\mu}^{\sigma} &=
\begin{cases}
\pm 1 \quad\; &\text{ if } \mu = |S_i\bar{\sigma}|\\
0 \quad\;&\text{ otherwise}
\end{cases}\\
\beta_{\mu}^{\sigma} &= 
\begin{cases}
\frac{f_{i\sigma}}{|S|} \alpha_{\mu}^{\sigma} &\text{ if } \mu = |S_i\bar{\sigma}|\\
0 &\text{ otherwise}
\end{cases}
\end{align*}
with $|\nu|\in\mathcal{P}^{\otimes n}$ denoting the Pauli group element $\nu\in\mathcal{P}_n$ with trivial prefactor.
\end{lem}
\vspace{12pt}
\begin{proof}
From the expression
\begin{align*}
E_\sigma =\frac{1}{2|S|} \sum_i S_i\bar{\sigma}
\end{align*}
it is clear that $\alpha_\mu^\sigma=0$ if $\mu\ne|S_i\bar{\sigma}|$ for all stabilizers $S_i$.

Let's now consider the case $\mu= |S_i \bar{\sigma}|$. First it is important to observe that the stabilizer $S_i$ and the Pauli operator $\sigma$ are uniquely determined by the product $|S_i \bar{\sigma}|$.

Since stabilizers and logical operators are hermitian 
and commute, $S_i\bar{\sigma}$ is hermitian, too. As a consequence each summand in the encoding operator $E_\sigma$ equals a distinct tensor product of Pauli operators with prefactor $\pm \frac{1}{2^n}$ and thus $\alpha_{|S_i \bar{\sigma}|}^{\sigma} =\pm 1$.

The expression for $\beta_{\mu}^{\sigma}$ follows directly from the coefficient comparision of $E_\sigma$ and
\begin{align*}
D_\sigma &= \frac{1}{|S|} \sum_i f_{i\sigma} S_i \bar{\sigma}\text{.}\qedhere
\end{align*}
\end{proof}
\vspace{12pt}

The above properties of the $\alpha$ and $\beta$ coefficients allow an easy computation of the coding map based on the expression provided by proposition \ref{prop:codingMap}. The expression gets even simpler if the original superoperator is diagonal in the Stokes representation, as we see in the next proposition.
The expression originates from \cite{rahn} and shows that concatenated stabilizer codes preserve the diagonality of superoperators:
\vspace{12pt}
\begin{prop}\label{thm:channelMapStab}
A stabilizer code $C(S)$ transforms a trace-preserving superoperator $T=[x,y,z]$ into the effective trace-preserving 
superoperator $\tilde{T} = \Omega^C(T)$ with
\begin{align*}
\tilde{T}_{\sigma\sigma'} = \delta_{\sigma\sigma'} \frac{1}{|S|} \sum_{S_i\in S} f_{i\sigma} x^{w_X(S_i\bar{\sigma})} y^{w_Y(S_i\bar{\sigma})} z^{w_Z(S_i\bar{\sigma})}
\end{align*} 
and $w_\nu(\mu)$ denoting the number of occurence (the weigth) of the Pauli operator $\nu\in\mathcal{P}$ in the tensor product $\mu\in \mathcal{P}_n$.
\end{prop}
\vspace{12pt}
\begin{proof}
Due to the diagonality of $T$ the statement of proposition \ref{prop:codingMap} simplifies to
\begin{align*}
\tilde{T}_{\sigma\sigma'} &= \sum_{\mu\in\mathcal{P}^{\otimes n}} \beta_{\mu}^\sigma \alpha_{\mu}^{\sigma'} \prod_{j=1}^n T_{\mu_j \mu_j}\\
&= \sum_{\mu\in\mathcal{P}^{\otimes n}}\beta_{\mu}^\sigma \alpha_{\mu}^{\sigma'} x^{w_X(\mu)} y^{w_Y(\mu)} z^{w_Z(\mu)}\text{.}
\end{align*}
The desired statement follows by replacing the $\beta$ coefficients with the expression of lemma \ref{lem:coefficients} allowing us to restrict the sum to operators 
$\mu=|S_i \bar{\sigma}|$.
\end{proof}
\vspace{12pt}

\section{Convergence of diagonal noise} 

The motivation for studying concatenated coding is the idea that the effective channel gets less noisy with every coding level and eventually converges to the identity. But of course this will only happen if both the code and the initial noise fulfill certain minimum requirements. Mathematically speaking the question is if the identity is an attracting fixed point of the coding map and if the initial noise lies within its basin of attraction.

In this section we will restrict our studies to the convergence of diagonal, trace-preserving superoperators under diagonality-preserving codes. For such superoperators the coding map is fully described by its \emph{diagonal-reduced coding map} variant 
\begin{align*}
&\Omega_d^C:\,\mathbb{R}^3\rightarrow\mathbb{R}^3\\
&\Omega_d^C\big([T_{XX},T_{YY},T_{ZZ}]\big):= [\tilde{T}_{XX},\tilde{T}_{YY},\tilde{T}_{ZZ}]
\end{align*}
with $\tilde{T}=\Omega^C(T)$.

\vspace{12pt}
\begin{definition}
Let $f: \mathbb{R}^k \rightarrow \mathbb{R}^k$.
\begin{itemize}
\item A point $p=f(p)$ is called a \emph{fixed point} of $f$.
\item The \emph{orbit} of a point $x\in \mathbb{R}^k$ is the series $\big(x,f(x), f^2(x), f^3(x), \dots\big)$ with $f^n$ indicating the $n$-times concatenation of $f$.
\item A fixed point $p$ is called \emph{locally attracting} iff it has a neighborhood $U\subseteq \mathbb{R}^k$ such that the orbit of every $x\in U$ converges towards $p$.
\item The \emph{basin of attraction} $B(p)$ of an attracting fixed point $p$ is its largest neighborhood such that the orbit of every point in $B(p)$ converges towards $p$.
\end{itemize}
\end{definition}
\vspace{12pt}

We consider an arbitrary norm $\Vert \cdot \Vert$ on $\mathbb{R}^k$ and write $\Vert Df(p)\Vert$ to denote the induced norm of the Jacobian $Df$ of a differentiable function $f: \mathbb{R}^k \rightarrow \mathbb{R}^k$ in the point $p$:
\begin{align*}
\big\Vert Df(p)\big\Vert = \sup\Big\{ \big\Vert \big(D f(p)\big) r\big\Vert \,\Big\vert\, r\in\mathbb{R}^k, \Vert r\Vert \le 1\Big\}\text{.}
\end{align*}

Proposition \ref{prop:codingMap} tells us that the diagonal-reduced coding map maps to a vector of polynomials and is therefore a $C_1$-function, i.e., its 
Jacobian exists and is continuous. This allows us to apply a standard result from analysis giving a sufficient condition for a fixed point $p$ to be attracting.
\vspace{12pt}
\begin{lem}\label{lem:attracting}
Let $f:\mathbb{R}^k\rightarrow \mathbb{R}^k$ be a $C_1$-function with fixed point $p$. If $\lambda_0:=\Vert D f(p)\Vert <1$ then $p$ is locally attracting.
\end{lem}
\vspace{12pt}
\begin{proof}
Choose a $\lambda$ with $\lambda_0<\lambda<1$. Since $Df(x)$ is continuous there exists a neighbourhood $U$ of $p$ such that $\Vert Df(x)\Vert \le \lambda$ for all $x\in U$. By the mean value theorem the orbits of all $x\in U$ converge to $p=f^n(p)$:
\begin{align*}
\lim_{n\rightarrow \infty}\Vert f^n(x) - f^n(p)\Vert &\le \lim_{n\rightarrow \infty} \lambda^n \Vert x - p \Vert = 0\text{.}\qedhere
\end{align*}
\end{proof}
\vspace{12pt}

The next proposition corresponding to \cite[theorem 3.3]{fern} shows that $\Vert D\Omega^C_d\big([1,1,1]\big)\Vert = 0$ for codes with a certain minimum error-correction capability. With the above lemma this indeed reveals 
 the identity channel as an attracting fixed point of the diagonal-reduced coding map.

\vspace{12pt}
\begin{prop}\label{thm:Jacobian0}
Let $C$ be a diagonality-preserving quantum code correcting any single qubit error. Then the Jacobian of the diagonal-reduced coding map $\Omega^C_d$ vanishes at $[1,1,1]$, exposing it as an attracting fixed point. 
\end{prop}
\vspace{12pt}
\begin{proof}
Consider the single qubit noise $T=[1,1-2\epsilon,1-2\epsilon]$ which imposes an $X$-error with probability $\epsilon$ and no error otherwise according to lemma \ref{lem:pauliChannel}. Since $C$ can correct any single qubit error the probability that a coding level on the product noise $T^{\otimes n}$ returns the identity channel is $1-\mathcal{O}(\epsilon^2)$. 
Hence
\begin{align*}
\Omega^C_d(T)=[1-\mathcal{O}(\epsilon^2),1-\mathcal{O}(\epsilon^2),1-\mathcal{O}(\epsilon^2)]\text{.}
\end{align*}

Because the effective channel does not contain a first order term in $\epsilon$, the derivative in the direction of $v_X=[0,-2,-2]$ vanishes:
\begin{align*}
D\Omega^C_d\big([1,1,1]\big)\cdot v_x &= \frac{d}{d\epsilon} \Omega^C_d\big( \underbrace{[1,1,1] + \epsilon v_X}_{=T}\big)_{\vert \epsilon=0}\\
&=[0,0,0]\text{.}
\end{align*}

The argument runs analogously for $v_Y=[-2,0,-2]$ and $v_Z=[-2,-2,0]$. As $v_X$, $v_Y$, $v_Z$ are linearly independent it follows that $D\Omega^C_d\big([1,1,1]\big)=0$.
\end{proof}
\vspace{12pt}

Notice that the above proposition only states that small diagonal noise will converge towards the identity by suitable concatenated coding. This can be concluded from the derivative of the restricted function $\Omega^C_d$ because the starting point is assumed to be diagonal and diagonality of the channel series is preserved.
It is important to stress that so far the proposition does not allow any conclusion for non-diagonal noise. In such a case the derivative of the full coding map $\Omega^C$ has to be checked instead. It is neither clear that this full derivative vanishes for the identity channel nor that the norm is even smaller than $1$. But in the next section we will at least prove that the identity is always an attracting fixed point of the full coding map for stabilizer codes of distance $\ge 3$.

\section{Convergence of general noise} 


In this section we extend our convergence study of concatenated coding from the previous section to non-diagonal noise but under the restriction of stabilizer codes. Stabilizer codes have the advantage that noise channels with small off-diagonal terms almost behave like diagonal channels, as shown by the next proposition, adapted from \cite[theorem 5.5]{fern}.

\vspace{12pt}
\begin{prop}\label{thm:differenceMap}
Consider coding with an $[n,1,d,w]$ stabilizer code against a quantum channel $T= D + \epsilon N$, $\left\vert N_{\sigma\sigma'}\right\vert\le 1$ for all $\sigma, \sigma'\in\mathcal{P}$, with $D$ containing the diagonal and $\epsilon N$ the off-diagonal elements in Stokes representation.
Then
\begin{align*}
\left\vert \big(\Omega^C(T)-\Omega^C(D)\big)_{\sigma\sigma'} \right\vert &\le
\begin{cases}
c_N \epsilon^d \quad&\text{ if }\sigma\ne\sigma'\\
c_N\epsilon^w\quad&\text{ if }\sigma=\sigma'
\end{cases}
\end{align*}
with
\begin{align*}
2^m\le c_N:=2^m \max_{\sigma\in\mathcal{P}}\sum_{S_i\in S} \left\vert \beta^\sigma_{\vert S_i \bar{\sigma}\vert}\right\vert\le 4^m\text{.}
\end{align*}
\end{prop}
\vspace{12pt}
\begin{proof}
For the off-diagonal elements we obtain, with the help of proposition \ref{prop:codingMap} and lemma \ref{lem:coefficients},
\begin{align*}
\left\vert \big(\Omega^C(T)\big)_{\sigma\sigma'}\right\vert
&
{=}  \left\vert \sum_{\mu,\nu\in\mathcal{P}^{\otimes n}} \beta_{\nu}^\sigma \alpha_{\mu}^{\sigma'} \prod_{k=1}^n T_{\nu_k \mu_k} \right\vert\\
&
\le  \sum_{S_i,S_j \in S}  \left\vert \beta^\sigma_{\vert S_i\bar{\sigma}\vert} \right\vert 
\left\vert \prod_{k=1}^n   T_{\vert S_i\bar{\sigma}\vert_k \vert S_j\bar{\sigma}'\vert_k} \right\vert\\
&\le 2^m \sum_{S_i\in S} \left\vert \beta^\sigma_{\vert S_i \bar{\sigma}\vert}\right\vert \epsilon^d\\
&\le c_N \epsilon^d\text{.}
\end{align*}
The third line follows from the second line since $\vert S_i \bar{\sigma}\vert$ and $\vert S_j \bar{\sigma}'\vert$ are multiplicatively related by an element from $N(S)\backslash S^\pm$ and hence differ in at least $d$ qubit positions. Thus the product contains at least $d$ off-diagonal matrix entries of $T$ while the remaining factors are upper-bounded by $1$.

For the diagonal elements we first derive analogously
\begin{align*}
&\left\vert \big(\Omega^C(T)-\Omega^C(D)\big)_{\sigma\sigma} \right\vert\\
&\quad
{=}  \left\vert \sum_{\mu,\nu\in\mathcal{P}^{\otimes n}} \beta_{\nu}^\sigma \alpha_{\mu}^\sigma
\left( \prod_{k=1}^n T_{\nu_k \mu_k} -  \prod_{k=1}^n D_{\nu_k \mu_k}  \right) \right\vert\\
&\quad
\le  \sum_{S_i,S_j \in S}  \left\vert \beta^\sigma_{\vert S_i\bar{\sigma}\vert} \right\vert 
\left\vert \prod_{k=1}^n   T_{\vert S_i\bar{\sigma}\vert_k \vert S_j\bar{\sigma}\vert_k}- \prod_{k=1}^n D_{\vert S_i\bar{\sigma}\vert_k \vert S_j\bar{\sigma}\vert_k} \right\vert\text{.}
\end{align*}
Since the two products are the same and cancel if $S_i=S_j$, we can restrict the sum to $S_i\ne S_j$. Given this condition the second product vanishes since it contains at least one off-diagonal factor. Recalling finally that $S_i\bar{\sigma}$ and $S_j\bar{\sigma}$ are multiplicatively related by a non-identity stabilizer whose weight is greater or equal to $w$, it follows that
\begin{align*}
&\left\vert \big(\Omega^C(T)-\Omega^C(D)\big)_{\sigma\sigma} \right\vert\\
&\qquad\le \sum_{S_i \ne S_j \in S} \left\vert\beta^\sigma_{\vert S_i\bar{\sigma}\vert}\right\vert \left\vert\prod_{k=1}^n T_{\vert S_i\bar{\sigma}\vert_k \vert S_j\bar{\sigma}\vert_k}\right\vert\\
&\qquad\le 2^m \sum_{S_i\in S} \left\vert \beta^\sigma_{(S_i \bar{\sigma})}\right\vert \epsilon^w\\
&\qquad\le c_N \epsilon^w\text{.}
\end{align*}

It remains to prove the bounds on the constant $c_N$. Lemma \ref{lem:coefficients} shows $\vert\beta^\sigma_{\left\vert S_i \bar{\sigma}\right\vert}\vert \le 1$ for all $S_i\in S$ and $\sigma\in\mathcal{P}$. Together with $\left\vert \beta^\mathbb{I}_\mathbb{I}\right\vert=1$, it follows directly that
\begin{align*}
2^m\le c_N &=2^m \max_{\sigma\in\mathcal{P}}\sum_{S_i\in S} \left\vert \beta^\sigma_{\left\vert S_i \bar{\sigma}\right\vert}\right\vert\le 4^m\text{.}\qedhere
\end{align*}
\end{proof}
\vspace{12pt}

The previous proposition tells us how a non-diagonal noise channel converges towards some diagonal channel under suitable concatenated stabilizer coding. But we are interested in the convergence towards a specific diagonal channel, namely the identity. For the off-diagonal elements, the previous proposition already provides a bound, but for the diagonal elements we have to combine the above result via the triangle inequality with the statement of proposition \ref{thm:Jacobian0}:

\vspace{12pt}
\begin{prop}\label{thm:errorScaling}
Consider coding with an $[n,1,d,w]$ stabilizer code with distance $d\ge 3$ and $w\ge 2$ against a quantum channel $T$ obeying $|(T-\Id)_{\sigma\sigma'}|\le\epsilon\le 1$ for all $\sigma,\sigma'\in\mathcal{P}$. Then there exists a constant $c$ such that
\begin{align*}
\left\vert\big(\Omega^C(T)- \Id \big)_{\sigma\sigma'}\right\vert \le c \epsilon^2 \qquad\forall\sigma,\sigma'\in\mathcal{P}\text{.}
\end{align*}
\end{prop}
\vspace{12pt}
\begin{proof}
Due to proposition \ref{thm:differenceMap} and the triviality of the case $\sigma=\sigma'=\mathbb{I}$, it only remains to show the statement for the cases $\sigma=\sigma'\in\{X,Y,Z\}$. For this write
\begin{align*}
T= \Id + \epsilon M + \epsilon N
\end{align*}
with $D:=\Id + \epsilon M$ just containing the diagonal and $\epsilon N$ the off-diagonal entries.

From proposition \ref{thm:Jacobian0} we know that the Jacobian of the diagonal-reduced coding map $\Omega^C_d$ vanishes at the point $[1,1,1]$ and that hence the Taylor series expansion of $(\Omega^C_d)_\sigma$ around this point does not contain a linear term. Since the Taylor series is finite and $\epsilon\le 1$, there exists a constant $c_M$ such that 
\begin{align*}
\left\vert\big(\Omega^C(D)-\Id\big)_{\sigma\sigma}\right\vert =\left\vert(\Omega^C_d)_\sigma(D)-1\right\vert \le c_M \epsilon^2\text{.}
\end{align*}
The triangle inequality and proposition \ref{thm:differenceMap} lead to
\begin{align*}
&\left\vert\big(\Omega^C(T)- \Id \big)_{\sigma\sigma}\right\vert\\
&\qquad\le \left\vert\big(\Omega^C(T)- \Omega^C(D)\big)_{\sigma\sigma}\right\vert + \left\vert\big(\Omega^C(D) - \Id \big)_{\sigma\sigma}\right\vert\\
&\qquad\le (c_N+c_M) \epsilon^2\text{.}\qedhere
\end{align*}
\end{proof}
\vspace{12pt}


So far we derived an expression for how noise decreases by one level of coding. Complete induction helps us to derive a non-recursive bound on the noise that remains after $k$ levels of concatenated coding:
\vspace{12pt}
\begin{lem}\label{lem:seriesAbsolute}
The recursive series $\epsilon_{k+1}=\alpha \epsilon_k^2$ has the explicit form
\begin{align*}
\epsilon_k = \frac{1}{\alpha} (\alpha \epsilon_0)^{2^k}\text{.}
\end{align*}
\end{lem}
\vspace{12pt}
\begin{proof}
The statement is proven by complete induction. For $k=0$ the statement is obviously fulfilled. Now assume the statement is true for a specific $k$. Then
\begin{align*}
\epsilon_{k+1} &= \alpha \epsilon_k^2\\
&\stackrel{\text{i.h.}}{=} \alpha \left(\frac{1}{\alpha} (\alpha \epsilon_0)^{2^k}\right)^2\\
&= \frac{1}{\alpha} (\alpha \epsilon_0)^{2^{k+1}}\text{.}\qedhere
\end{align*}
\end{proof}
\vspace{12pt}
\begin{cor}\label{cor:absoluteConvergence}
Consider the setting of proposition \ref{thm:errorScaling} and denote by $T^{(k)}$ the effective channel after $k$ levels of concatenated coding against the noise channel $T=T^{(0)}$. Then
\begin{align*}
\left\vert\big(T^{(k)}- \Id \big)_{\sigma\sigma'}\right\vert \le \frac{1}{c} (c \epsilon)^{2^k} \qquad\forall\sigma,\sigma'\in\mathcal{P}\text{.}
\end{align*}
\end{cor}
\vspace{12pt}
\section{Robustness of $\QMA$ against constant qubit noise}

In this section we combine the previous results to prove that $\QMA_T=\QMA$ even holds for channels $T$ of constant noise. The central idea is to verify a $\QMA$-problem receiving a disturbed encoded version of the witness that the undisturbed $\QMA$-protocol would receive. The $\QMA_T$-protocol carries out error correction and decoding to extract the original witness and then carries out the original $\QMA$-protocol. Despite the efficiency restriction to maximally polylogarithmic many levels of coding the extracted witness will be sufficiently close to the original witness to guarantee any acceptance probability in the usual range.

The main theorem is preceded by the following lemma allowing us to convert between the bound on the diamond norm of $T-\Id$ and bounds on the matrix entries in Stokes representation:
\vspace{12pt}
\begin{lem}\label{lem:normsTranslation}
For all superoperators $T$ 
and all $\sigma,\sigma'\in\mathcal{P}$ it follows that
\begin{align*}
\frac{1}{16c_\Diamond} \Vert T-\Id\Vert_\Diamond \le \vert(T-\Id)_{\sigma\sigma'}\vert \le \Vert T-Id\Vert_\Diamond
\end{align*}
with $c_\Diamond$ the maximum diamond norm over all superoperators whose matrix entries in Stokes representation all vanish except from one equaling $1$.
\end{lem}
\vspace{12pt}
\begin{proof}
The first inequality is derived simply by applying the triangle inequality to the 16 matrix entries of $T-\Id$ in Stokes representation. Observe, that for trace-preserving $T$ the number 16 can be replaced by 12 and that the constant $c_\Diamond$ can be easily computed via semi-definite programming \cite{watrousSDP}. For our purposes the actual value of the positive constant is not important. Of course, $c_\Diamond\ne 0$, since $\Vert\cdot\Vert_\Diamond$ is a valid norm. 

The second inequality follows from
\begin{align*}
\Vert T-\Id\Vert_\Diamond
&\ge \Vert T-\Id\Vert_1\\
&\ge \frac{1}{2}\Vert (T-\Id)(\sigma')\Vert_1\\
&= \frac{1}{2} \left\Vert \sum_{\sigma''\in\mathcal{P}} (T-\Id)_{\sigma''\sigma'} \sigma''\right\Vert_1\\
&=\frac{1}{2} \max_{-\mathbb{I}\le\Lambda\le \mathbb{I}} \tr\left[\Lambda \left(\sum_{\sigma''\in\mathcal{P}} (T-\Id)_{\sigma''\sigma'} \sigma'' \right) \right]\\
&\ge\vert(T-\Id)_{\sigma\sigma'}\vert
\end{align*}
with the last two lines following from the equality
\begin{align*}
\Vert \omega\Vert_1=\max_{-\mathbb{I}\le\Lambda\le \mathbb{I}} \tr\left[\Lambda\omega\right]
\end{align*}
for any hermitian operator $\omega$ \cite[Exercise 9.1.4]{wilde} and the fact that the maximizing set comprises the operators $\Lambda=\pm \sigma$.
\end{proof}
\vspace{12pt}
\begin{thm}\label{thm:main}
There is a constant $\epsilon_\Diamond>0$ such that $\QMA_T=\QMA$ for every quantum channel $T$ with $\Vert T-\Id\Vert_\Diamond\le\epsilon_\Diamond$.

Furthermore, for a channel $T$ of the above form, $\QMA_T(c,s)=\QMA_T$ for all polynomial-time computable functions $c$ and $s$ between $e^{-q}\le s, c\le 1-e^{-q}$ with gap $c-s\ge 1/q$ and $q\in\poly(n_\text{in})$.
\end{thm}
\vspace{12pt}
\begin{proof}
To prove the equality statement and amplification at the same time, will show that $\QMA\subseteq\QMA_T(1-e^{-r},e^{-r})$ for arbitrary $r\in\poly(n_\text{in})$.

Assume the setting of concatenated coding against the noise channel $T$ as in proposition \ref{thm:errorScaling} and corollary \ref{cor:absoluteConvergence}. Since $|(T-\Id)_{\sigma\sigma'}|\le \epsilon_\Diamond$ according to lemma \ref{lem:normsTranslation}, choosing $\epsilon_\Diamond  < \min\{1,c^{-1}\}$ ensures that the matrix entries decrease superexponentially with the number of coding levels $k$:
\begin{align*}
|(T^{(k)}-\Id)_{\sigma\sigma'}|\le \frac{1}{c} \left(c \epsilon_\Diamond \right)^{2^k} \qquad\forall\sigma,\sigma'\in\mathcal{P}\text{.}
\end{align*}

Now let $A$ be a problem in $\QMA$ and $V$ a $\QMA(1-\frac{1}{2}e^{-r},e^{-r})$-verifier for it. Let $V'$ be the verifier $V$ preceded by $k$ levels of error correction and decoding. Analogously to the proof of theorem \ref{thm:QMA_T=QMA}, the acceptance probability $c'$ of $V'$ in the completeness case when receiving the disturbed $k$-times encoded witness is lower bounded by
\begin{align*}
c' &\ge 1-\frac{1}{2}e^{-r} -\frac{1}{2}n_\text{w} \Vert T^{(k)} -\Id\Vert_\Diamond\\
&\ge 1-\frac{1}{2}e^{-r}- \frac{8c_\Diamond}{c} n_\text{w} \left( c \epsilon_\Diamond\right)^{2^k}
\end{align*}
with $n_\text{w}\in\poly(n_\text{in})$ denoting the length of the original undisturbed witness.

Clearly $k:=\log s$ with $s\in\poly(n_\text{in})$ can be choosen such that $c'\ge 1-e^{-r}$ for large enough input lengths $n_\text{in}$. For smaller input lengths let $V'$ give the correct output by hard encoding. Since each of the 
\begin{align*}
n_\text{w} \sum_{i=0}^{k-1} n^i \in\poly(n_\text{in})
\end{align*}
error correction and decoding operations acts on a finite size Hilbert space, the overall error correction and decoding can be accomplished efficiently and $V'$ is a valid $\QMA_T(1-e^{-r},e^{-r})$-verifier for $A$.
\end{proof}
\vspace{12pt}

We will spend the next two theorems on actually computing a noise value for which $\QMA_T=\QMA$ holds. The first theorem states the value for general channels and the next theorem derives the value for the partly depolarizing and the partly dephasing channel, which are of special interest since they lead to the complexity classes $\BQP=\QMA_{T^\text{depol}_1}$ and $\QCMA=\QMA_{T^\text{deph}_1}$ for the highest error parameter $\epsilon=1$.

Rewritten in Stokes representation the partly depolarizing and dephasing channel are diagonal and equal
\begin{align*}
T^\text{depol}_\epsilon&=
[1-\epsilon,1-\epsilon,1-\epsilon]\\
T^\text{deph}_\epsilon&=
[1,1-\epsilon,1-\epsilon]\text{.}
\end{align*}

We do not claim to present the highest error values that can be tolerated by the constructed scheme of concatenated coding since we just compared the performances of the 5-qubit, the Steane 7-qubit and the Shor code for the proofs of the following theorems. The coding maps of these codes possess the necessary superexponential convergence towards the identity channel since they are $[n,1,d,w]$ stabilizer codes with $d\ge 3$ and $w\ge 2$. The computation of the tolerable error is accomplished by specifying the 
basin of attraction of the identity channel.
Thanks to the structure of the codes, some of the diagonal coding map entries separate from the others. This reduces the task to determining the lower boundary of the basin of attraction of the fixed point $1$ for a one-dimensional function which is just given by the next lower fixed point.
\vspace{12pt}
\begin{thm}\label{thm:values}
$\QMA_T=\QMA$ if $T$ is diagonal in Stokes representation and $\Vert T-\Id\Vert_\Diamond\le 0.18$ or if $T$ is arbitrary and $\Vert T-\Id\vert_\Diamond \le 0.014$.
\end{thm}
\vspace{12pt}
\begin{proof}
The values can be derived using the 5-qubit code as specified in table \ref{tab:codes}. The diagonal-reduced coding map of the 5-qubit code is
\begin{align*}
\Omega^C_d([x,y,z])&=[f(x,y,z),f(y,z,x),f(z,x,y)]\\
f(x,y,z)&=-\frac{1}{4}x^5 + \frac{5}{4}xy^2 +\frac{5}{4}xz^2-\frac{5}{4}xy^2z^2\text{.}
\end{align*}
For the partly depolarizing channel with $x=y=z$ the function $f$ simpifies to the one-dimensional function
\begin{align*}
f_{1D}(x):=f(x,x,x)=\frac{5}{2}x^3-\frac{3}{2}x^5
\end{align*}
whose next fixed point below $1$ turns out to equal $\sqrt{2/3}$. Hence the depolarizing channel $T_\epsilon^\text{depol}$ with $\epsilon < 1- \sqrt{2/3}<0.18$ will converge towards the identity under concatenated 5-qubit codes, proving $\QMA_{T_\epsilon^\text{depol}}=\QMA$  via theorem \ref{thm:central}.

The function $f$ guarantees that every diagonal, trace-preserving superoperator $T$ with elements within the interval ${(\sqrt{2/3},1]}$ fulfills
\begin{align*}
(\Omega^C(T^\text{depol}_\epsilon)_{\sigma\sigma} \le (\Omega^C(T))_{\sigma\sigma}\le 1 \qquad\forall\sigma\in\mathcal{P}\text{.}
\end{align*}
Hence every diagonal channel $T$ with $\vert (T-\Id)_{\sigma\sigma}\vert <1-\sqrt{2/3}$ shows a sufficient convergence for the proof of theorem \ref{thm:central}. This is also true for every non-diagonal channel $T$ with $\vert (T-\Id)_{\sigma\sigma}\vert < (c_N+c_M)^{-1}$ according to proposition \ref{thm:differenceMap}. The values can be computed with $c_M=(1-\sqrt{2/3})^{-1}$, $c_N=64$ and $\vert (T-\Id)_{\sigma\sigma}\vert\le \Vert T-\Id\Vert_\Diamond$.
\end{proof}
\vspace{12pt}
\begin{thm}
\begin{align*}
\QMA_{T_\epsilon^\text{depol}} &=
\begin{cases}
\QMA &\text{ for } \epsilon\le 0.18\\
\BQP \quad&\text{ for } \epsilon=1
\end{cases}\\
\QMA_{T_\epsilon^\text{deph}} &=
\begin{cases}
\QMA &\text{ for } \epsilon\le 0.27\\
\QCMA &\text{ for } \epsilon=1\text{.}
\end{cases}
\end{align*}
\end{thm}
\vspace{12pt}
\begin{figure}[bt]
\begin{minipage}[h]{0.49\linewidth}\centering
\includegraphics[width=\textwidth]{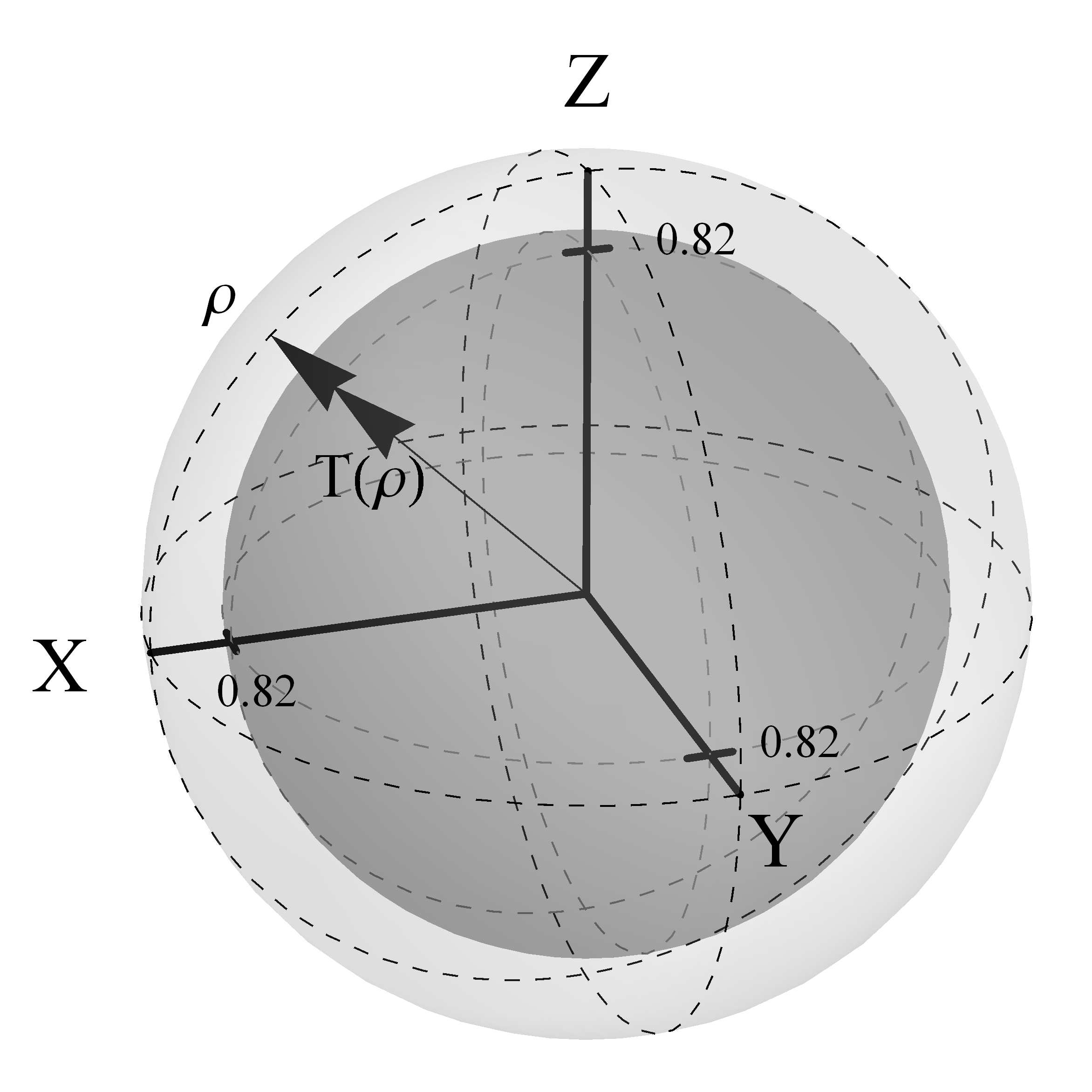}
\end{minipage}
\begin{minipage}[h]{0.49\linewidth}\centering
 \includegraphics[width=\textwidth]{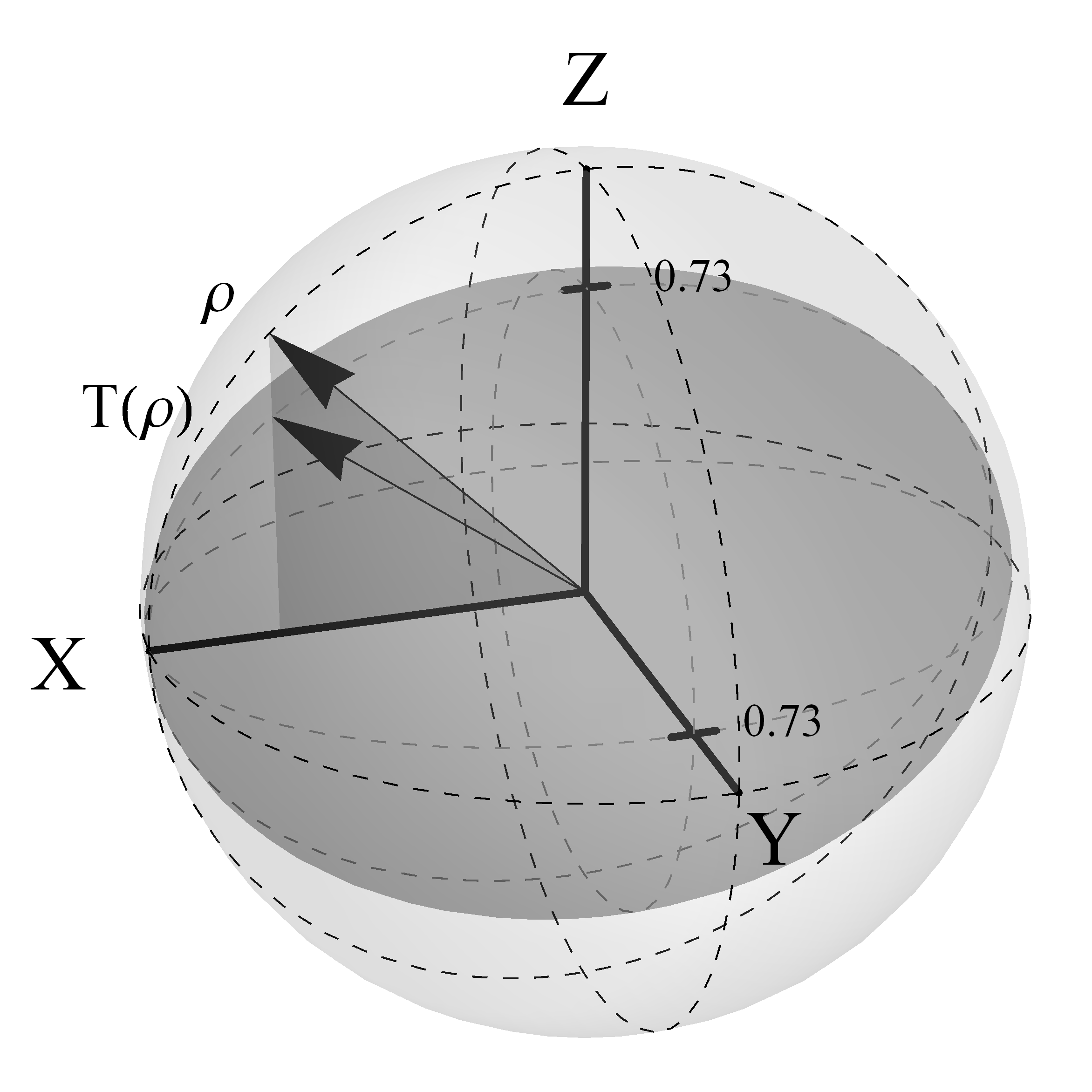}
\end{minipage}
\fcaption{\label{fig:bloch}$\QMA$ stays unchanged even if each witness qubit is 18\% depolarized (left) or 27\% dephased (right).}
\end{figure}
\begin{proof}
The bound on $\epsilon$ guaranteeing $\QMA_{T_\epsilon^\text{depol}}=\QMA$ has already been derived via the 5-qubit code in the proof of theorem \ref{thm:values}.

The higher bound on $\epsilon$ guaranteeing $\QMA_{T_\epsilon^\text{deph}}=\QMA$ can be derived by the Shor code as specified in table \ref{tab:codes}. Its diagonal-reduced coding map is of the form
\begin{align*}
\Omega^C_d([x,y,z])&=[f(x),
g(x,y,z),
h(z)]\\
h(z)&=\left(\frac{3z}{2}- \frac{z^3}{2}\right)^3\text{.}
\end{align*}
The next fixed point below $1$ of the function $h$ is $<0.73$. Furthermore we already start at the attractive point $1$ of the function $f$ with the dephasing channel $T_\epsilon^\text{deph}=[1,1-\epsilon,1-\epsilon]$.
By lemma \ref{lem:pauliChannel} the convergence of two diagonal channel elements towards $1$ is a sufficient criterion for all three converging towards $1$. Hence the channel $T_\epsilon^\text{deph}$ with $\epsilon<0.27$ will converge towards the identity under concatenated Shor codes, proving $\QMA_{T_\epsilon^\text{deph}}=\QMA$ via theorem \ref{thm:central}.
\end{proof}
\vspace{12pt}

\section{Open questions}

The result of theorem \ref{thm:central} can also be interpreted in the way that every $\QMA$-problem has a verifying protocol such that the witness supplied in the completeness case has the form of a certain channel output. One may now consider again encoding this witness and exposing it to noise. Theorem \ref{thm:central} promises again that there exists a $\QMA$-verifier receiving witnesses of this form. 
This procedure is unlikely to lead to a higher error robustness, but
\vspace{12pt}
\begin{ques}
Is there a combination of a specific code and noise that leads to an intriguing witness structure if interleaved constantly many times?
\end{ques}
\vspace{12pt}

\begin{ques}
Which interesting cases occur with an adaption of the disturbed witness scenario to other complexity classes such as $\QIP$, $\QAM$, $\QMAM$ \dots \cite{kitaevQIP, marriottWatrous}?
\end{ques}
\vspace{12pt}
Notice that we did not include restricted versions of $\QMA$ such as $\QCMA$ in the previous list, since we actually introduced the $\QMA_T$ framework to be able to describe these classes. Hence, it conforms with the concept of unification to consider $\QCMA$ rather as $\QMA_T$ with $T$ a fully dephasing channel.
\vspace{12pt}
\begin{ques}
Is there a compelling formulation of a quantum PCP conjecture \cite{aharonovVidickPCP} in terms of $\QMA_T$, e.g., $\QMA$ equals $\QMA_T$ with such a noise $T$ that the expected number of restored witness qubits is constant?
\end{ques}
\vspace{12pt}

\vspace{12pt}
\begin{ques}
To which value can the upper bound on the tolerable channel error be improved?
\end{ques}
\vspace{12pt}
\begin{ques}
Which lower bound on the error $\epsilon$ can be proven such that $\QMA_{T^\text{depol}_\epsilon}=\BQP$ and $\QMA_{T^\text{deph}_\epsilon}=\QCMA$?
\end{ques}
\vspace{12pt}
Quantum Shannon theory \cite{wilde,nielsen} might provide some statements helping to answer this question. However, this question exceeds a simple capacity argument, since the capacity of a channel $T$ (corresponding to the highest possible number of transmitted logical qubits per physical qubits in the limit for infinite message length) may well be $0$, but if the convergence is not faster than inverse polynomial, $\QMA_T$ may still have the power of $\QMA$.

\nonumsection{Acknowledgements}
\noindent
I thank Tobias J. Osborne for many helpful discussions and ideas.
This  work  was  supported  by  the  ERC grants QFTCMPS, and SIQS, and through the DFG by the cluster of excellence EXC 201 Quantum FQ Engineering and Space-Time Research, and the Research Training Group 1991.



\nonumsection{References}


\end{document}